\documentclass[aps,prd,superscriptaddress,longbibliography,amsmath,amssymb,twocolumn]{revtex4-2}

\usepackage[utf8]{inputenc}
\usepackage{graphicx}
\usepackage[amssymb,Gray]{SIunits}
\usepackage{textcomp}
\usepackage{physics}
\usepackage{calc}
\usepackage{tikz}
\usepackage{braket}
\usepackage{tensor}
\usepackage[export]{adjustbox}

\usepackage{hyperref}
\hypersetup{
    colorlinks=true,
    linkcolor=blue,
    filecolor=magenta,      
    urlcolor=cyan,
    pdftitle={Overleaf Example},
    pdfpagemode=FullScreen,
    }
\usepackage{xcolor}

\newcommand{\rmi}{\ensuremath{\mathrm{i}}}

\newcommand{\rmd}{\ensuremath{\mathrm{d}}}

\newcommand{\nnl}{\nonumber\\}
\renewcommand{\vec}[1]{\mathrm{\mathbf{#1}}}

\newcommand{\schr}{Schr\"odinger}
\newcommand{\dbbt}{de Broglie-Bohm theory}

\newsavebox\CBox
\newcommand\barletter[2][0.5pt]{%
  \ifmmode\sbox\CBox{$#2$}\else\sbox\CBox{#2}\fi%
  \makebox[0pt][l]{\usebox\CBox}%  
  \rule[0.5\ht\CBox-#1/2]{\wd\CBox}{#1}}

\begin{document}

%\preprint{}

\title{Entanglement generation \& dynamical equilibrium within \dbbt}


%\author{}
%\email[]{}
%\thanks{}
%\affiliation{}
%
\author{M.\ Kemal D\"oner}
\email[]{kemal.doner@uni-jena.de}
%\author{Andr\'e Gro{\ss}ardt}
%\email[]{andre.grossardt@uni-jena.de}
\affiliation{Institute for Theoretical Physics, Friedrich Schiller University Jena, Fr\"obelstieg 1, 07743 Jena, Germany}


\date{\today}

% PRL: abstract max. 600 characters
\begin{abstract}
    In this work, we reassess the hybrid semiclassical-gravity model introduced in~\cite{doner2022gravitational}, revealing its capacity to induce entanglement even from the \emph{limit of weak entanglement}. By incorporating feedback terms sourced from gravitational potential interactions between subsystems—built upon a refined TELB framework~\cite{norsen2010theory}—we demonstrate how these interactions alone can generate entanglement. Building on this insight, we propose a novel communication scheme and introduce the concept of dynamical equilibrium, which formalizes a stable strategy for subsystems. This framework draws inspiration from the quantum equilibrium hypothesis of de Broglie–Bohm theory~\cite{durr2009bohmian} and Bohm–Vigier’s causal interpretation~\cite{bohm1954model}.
\end{abstract}

%\keywords{}

\maketitle

% If in two-column mode, this environment will change to single-column
% format so that long equations can be displayed. Use sparingly.
%\begin{widetext}
%\end{widetext}

%%%%%%%%%%%%%%%%%%%%%%%%%%%%%%%%%%%%%%%%%%%%%%%%%%%%%%%%%%%%%%%%%%%%%%%
%%%  START  DOCUMENT  %%%%%%%%%%%%%%%%%%%%%%%%%%%%%%%%%%%%%%%%%%%%%%%%%
%%%%%%%%%%%%%%%%%%%%%%%%%%%%%%%%%%%%%%%%%%%%%%%%%%%%%%%%%%%%%%%%%%%%%%%

    \section{Introduction}
    %Lately, hybrid models on the subject of quantum gravity gained much interest due to their main aspect of \emph{bendability} for the selected limit that they are considered, and their versatility. Adding to that, they can also funded from many different approaches (or interpretations) of quantum theory, with that being said we propose that two of them gain a special consideration due to their intuitive idea and the potential connection between them. The first of these two models funded from the collapse theory (directly Ghirardi-Rimini-Weber theory) which is called as GRW model with massive flashes (GRWmf) that is introduced by Tilloy \cite{tilloy2018ghirardi}.
    Recently, a criticism~\cite{ward2025semiclass} has been raised that the Bohmian potential, $V_\text{bb}$, introduced in~\cite{doner2022gravitational} as an analogue of the Schrödinger–Newton equation (SNE)~\cite{bahrami2014schrodinger} (while neglecting self-interaction terms), cannot create entanglement due to two conditions: first, that it is additively separable, and second, that the single-particle wave equations have unique solutions $\psi_i \in \mathcal{H}$ for given wave functions.

    This critique, particularly regarding the first condition, is understandable given the resemblance of $V_\text{bb}$ to the semiclassical potential $V_\text{sc}$, except for the denominators—where in the former, Bohmian trajectories are deliberately utilized instead of the expectation value of the other particle's position, as is the case for the latter. (Additionally, for $V_\text{bb}$, the term function $\gamma$ introduces another distinction, that will be discussed later). Thus, it is clear to see why one might initially assume that these potentials yield the same outcome if they are unaware of the differences in these denominators. Meanwhile, it is worth noting that the weak entanglement limit was implicitly mentioned in the model to highlight these differences by considering the bare minimum—where the interaction potential dominates in comparison to the entanglement potential fields. 
    
    Regarding the second condition, it is clear that approximating $X_i(t) \approx u_i(t)$, leads to two fully decoupled \schr~equations for $\psi_i$ as shown in~\cite{doner2022gravitational}. Consequently, this condition will be considered resolved and neglected for the remainder of the article.
    
    Therefore, starting with the next section, we will focus on the first condition and introduce an assessment of our model’s ability to generate entanglement in the weak entanglement limit, where feedback terms developed from the interaction potentials will be responsible for the emergence of the vital tools: entanglement potential fields in the original model. 
    
    In the third section, the dynamic nature of these terms will be explored by introducing the concept of \emph{dynamical} which serves as a dynamical version of the quantum equilibrium hypothesis in \dbbt. This will be followed by the assumption that dynamical equilibrium acts as a stable strategy between the subsystems, where they interact with each other while aiming to conserve their wave function distribution during the interaction process. For the stable strategy, Nash’s theorem will be utilized and applied in numerical calculations, leading to a simulation of the trajectories of each subsystem.
    %In the fourth section, the connection between dynamical equilibrium and continuous stochastic measurement will be introduced. To achieve this, we will consider the stochastic version of Bohmian trajectories by exploring open quantum systems, where the environment of the focused subsystem is not limited to the \emph{other} subsystem but includes all surrounding factors. In this context, an equivalent of the Langevin equation will be introduced within the trajectory-based approach of \dbbt.
    
    In the fourth section, the conclusion of this article will be presented. It will contain a discussion of the findings together with future direction of them considering possible connections to relative topics.
	\section{Generating Entanglement}
	Consider two reference scenarios: a two-particle system that is initially separated and isolated, and a version of it without isolation. In first scenario,the total potential is clearly additively separable and can therefore be expressed as a sum. In contrast, in the second scenario, this separability no longer holds. Thus, when interactions are present, the total potential defined via $V_\text{bb}$ is neither additively separable nor does it lead to fully decoupled Schrödinger equations—unlike $V_\text{sc}$. This is a direct consequence of the use of conditional wave functions (CWFs) and Bohmian trajectories, as discussed in detail in~\cite{doner2022gravitational}.
	
	That being said, it should be acknowledged that the transition from first to second scenario was rather swift in the original work, with several details left implicit. Most notably, the role of interactions in generating entanglement potential fields—as introduced through the effective potential equations—was only briefly mentioned in the context of the weak entanglement limit, without a detailed derivation of the underlying process. We believe this gap warrants closer attention, and addressing it will be the focus of the following discussion.
	
	To that end, we consider a third scenario in which the entanglement potential fields are absent from the effective potentials, $V_i^\text{eff}$, for a certain period. This assumption is justified, as both orders of these fields are not $x$-dependent, as shown in~\cite{doner2022gravitational}. As a result, the dynamical evolution of particle 1’s conditional wave function (CWF) is governed by:
	\begin{align}\label{eqn:time-evo_CWF1-no-iso}
		\rmi \hbar \frac{\partial \psi_1(t,x)}{\partial t} = &-\frac{\hbar^2}{2m_1} \frac{\partial^2 \psi_1(t,x)}{\partial x^2} \nnl
		&+ V[t,x,X_2(t)]\psi_1(t,x)
	\end{align}
	In this regime, Eq.~\eqref{eqn:time-evo_CWF1-no-iso}, the only $x$-dependent term in $V_i^\text{eff}$ is the conditional potential $V[t,x,X_2(t)]$, for evolution of particle 1. In the absence of interaction, it reduces to $V_1(t,x)$, thus no entanglement is generated between the particles. However, once interaction occurs—which will be defined shortly—effects arise directly from $V[t,x,X_2(t)]$ (analogously for particle 2). 
	
\begin{figure}
		\centering
		\includegraphics[scale=0.55]{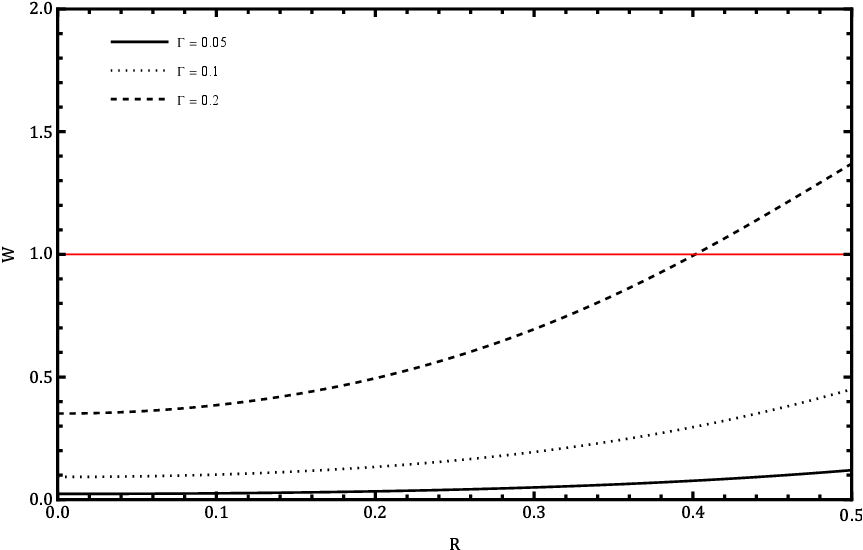}
		\caption[Entanglement witness $W$ as a function of $R$ for the wide wave function case due to weak entanglement]{Entanglement witness $W$ as a function of $R$ for the wide wave function case $\Delta x=0.25$, $\delta x=0.1$ and with shorter values for $\Gamma$ of $0.05, 0.1, 0.2$ due to weak entanglement.
		\label{fig:witness-gamma_short-plots}}
	\end{figure}

	In the case of initially separated particles with interactions present but before the emergence of any entanglement potential fields, again following~\cite{doner2022gravitational}, the effective potential; $V_i^\text{eff(isep)}$ (here \textit{isep} refers to "initially separated with emergent potential fields," describing the effective potential for a system of particles that start separated, with interactions present but without entangled potential fields having yet emerged) is defined as:
	\begin{subequations}\label{eqn:eff-pot-isep}
		\begin{align}
			V_1^\text{eff(isep)} &= \gamma_0(X_1,X_2) -\frac{G m_1 m_2}{\abs{X_1 - X_2}} -\frac{G m_1 m_2}{\abs{x - X_2}} \\
			V_2^\text{eff(isep)} &= \gamma_0(X_1,X_2) -\frac{G m_1 m_2}{\abs{X_1 - X_2}} -\frac{G m_1 m_2}{\abs{X_1 - x}}
		\end{align}
	\end{subequations}
	In~\cite{doner2022gravitational}, this situation handled by the choice of $\gamma_0$ (arbitrary function) cancelling relative conditional potential (second term) in Eq.~\eqref{eqn:eff-pot-isep} as a convenience coming from their negligibility compared to entanglement potential fields. However, in the current scenario, this cancellation is not valid; therefore, both potential terms in Eq.~\eqref{eqn:eff-pot-isep} should be retained (with $\gamma_0$ incorporated into $V[t,X_1(t), X_2(t)]$). Furthermore, in the absence of these fields, it follows from~\cite{norsen2010theory} that the only potential terms that remain are given by $P_n$, which accounts for the interactions through the gradient of these potentials, as defined in~\cite{norsen2010theory}. Therefore, the only surviving terms (for the first-order) are:
	\begin{subequations}
		\begin{align}\label{eqn:grad_con&relcon1}
			\frac{\partial V}{\partial x_2}[t,x,X_2(t)] &- \frac{\partial V}{\partial x_2}[t,X_1(t),X_2(t)] = \nnl 
			&\frac{\partial }{\partial x_2} \bigg[\frac{G m_1 m_2}{\abs{x - X_2}}\bigg] - \frac{\partial }{\partial x_2} \bigg[\frac{G m_1 m_2}{\abs{X_1 - X_2}}\bigg]\\
			\frac{\partial V}{\partial x_1}[t,X_1(t),x] &- \frac{\partial V}{\partial x_1}[t,X_1(t),X_2(t)] = \nnl
			&\frac{\partial }{\partial x_2} \bigg[\frac{G m_1 m_2}{\abs{X_1 - x}}\bigg] - \frac{\partial }{\partial x_1} \bigg[\frac{G m_1 m_2}{\abs{X_1 - X_2}}\bigg] \label{eqn:grad_con&relcon2}
		\end{align}
	\end{subequations}
	Eq.~\eqref{eqn:grad_con&relcon1} shows that if the gradient of the conditional potential, $V[t,x,X_2(t)]$, evaluated at $x_2 = X_2(t)$, differs from the gradient of the relative conditional potential, $V[t,X_1(t),X_2(t)]$, evaluated at $x_1 = X_1(t)$ and $x_2 = X_2(t)$—both derived from the initial potential $V[t,x_1,x_2]$—then the resulting non-zero difference can be regarded as the sole source of entanglement generation from an initially non-entangled state. A similar conclusion holds for Eq.~\eqref{eqn:grad_con&relcon2}. In other words, entanglement arises when the pilot waves (i.e., the CWFs) of the particles begin to overlap in physical space, as indicated by the non-vanishing values of Eqs.~\eqref{eqn:grad_con&relcon1} and~\eqref{eqn:grad_con&relcon2}. To address the emergence of entanglement, a modification—or more precisely, an alternative formulation—of the entanglement fields is required. This alternative formulation must ensure that the entanglement fields vanish in the absence of entanglement~\cite{norsen2010theory}; more accurately, it must guarantee this absence, (where Norsen calls it marginally improved TELB). This results in the following evolution equation:
	\begin{widetext}
		\begin{equation}\label{eqn:time-evo_CWF_one_withaltentfields-explicit}
			\begin{split}
				\rmi \hbar \frac{\partial \psi_1(t,x)}{\partial t} = &-\frac{\hbar^2}{2m_1} \frac{\partial^2 \psi_1(t,x)}{\partial x^2} + V[t,x,X_2(t)]\psi_1(t,x) + f(t) \psi_1(t,x)\\
				&+ \frac{\rmd X_2(t)}{\rmd t} \int_0^\tau \rmd t \bigg(\frac{\partial V}{\partial x_2}[t,x,X_2(t)] - \frac{\partial V}{\partial x_2}[t;X_1(t),X_2(t)] \bigg)\psi_1(t,x)\\
				&+ \frac{\rmi\hbar}{2m_2} \int_0^\tau \rmd t \bigg(\frac{\partial^2 V}{\partial x_2^2}[t,x,X_2(t)] - \frac{\partial^2 V}{\partial x_2^2}[t,X_1(t),X_2(t)] \bigg)\psi_1(t,x)\\
			\end{split}
		\end{equation}
	\end{widetext}	
	Eq.~\eqref{eqn:time-evo_CWF_one_withaltentfields-explicit} demonstrates how the pilot wave of particle 1 is dynamically influenced by weak entanglement, ultimately leading to the generation of entanglement. Comparing Eq.~\eqref{eqn:time-evo_CWF1-no-iso} with~\eqref{eqn:time-evo_CWF_one_withaltentfields-explicit} highlights not only the source of entanglement but also its mechanism, specifically through the overlap of the pilot waves of the particles via conditional and relative conditional potentials.
	In ~\cite{doner2022gravitational}, this is not explicitly presented but only mentioned by the discussion of the role of the $\gamma$ terms ($\gamma_0$ or $\gamma_R$), and stated that entanglement can be observed using phases, regardless of the choice of $\gamma_R$, simply by appropriately rescaling the parameter $\Gamma$ via the mass $m$ and flight time $\tau$. Therefore, following this statement and assuming shorter flight times, one can obtain the plot for the Witness function, $W$, of weak entanglement, as shown in Fig.~\eqref{fig:witness-gamma_short-plots}.
	The plot in Fig.~\eqref{fig:witness-gamma_short-plots} demonstrates a gradual increase in entanglement over time until it eventually surpasses the thin red line, which marks the onset of entanglement potential fields dominance. Consequently, entanglement is generated (or increased) from a state of weak entanglement. This analysis illustrates how entanglement evolves over time due to the interplay between gravitational potentials and their feedback on the particle’s pilot wave (CWF).
	
	\section{Dynamical Equilibrium}
	In the previous section, the model’s ability to generate entanglement, even in the absence of entanglement potential fields, through particle interactions demonstrated, as shown in Eq.~\eqref{eqn:time-evo_CWF_one_withaltentfields-explicit}. However, it was not explicitly addressed how the feedback terms operate during the evolution of pilot waves—specifically, how they manage the transition from effective wave functions (EWFs)~\cite{durr2009bohmian} to the CWFs and back to the EWFs. (This discussion implicitly considers scenarios where particles take the closest pathways in order to exhibit interactions, as in the Bose experiment~\cite{bose2017spin} or similarly~\cite{marletto2017gravitationally}.)
	
	To clarify this, we introduce concept of dynamical equilibrium, which is rooted in quantum equilibrium hypothesis—and its conditional version of \dbbt~\cite{durr2009bohmian}—and draws on the work of Bohm and Vigier namely causal interpretation~\cite{bohm1954model}. Briefly, this concept provides a framework for describing interactions between subsystems over a given time interval. During this interval, each subsystem’s pilot wave functions as a CWF, and the overlap of them in physical space is governed by the orders of feedback terms in Eq.~\eqref{eqn:time-evo_CWF_one_withaltentfields-explicit}.
	
	Furthermore, dynamical equilibrium defines a progression specifically centred around the conditional quantum equilibrium hypothesis. This progression can be characterized using the better-response function (see Appendix~[\ref{app:Nash-Th}]) from Nash’s theorem~\cite{nash1950equilibrium, jiang2009tutorial}, which determines the stable strategy adopted by the subsystems, Eq.~\eqref{eqn:each-com}. Notably, this function ensures continuous updates to each subsystem’s CWF, Eq.~\eqref{eqn:gain} as their distribution converges toward an equilibrium point (or set of equilibria). In the context of \dbbt, this equilibrium is defined by the conditional quantum equilibrium hypothesis (see Appendix~[\ref{app:best-response}]) for the subsystems, which serves as a reference point (where it accounts for causal interpretation's constant limit) for dynamical equilibrium. 
	
	In terms of stable strategy, the connection between Nash’s theorem and \dbbt~is encapsulated in the strategy vector $\varsigma$. For whole system, $\varsigma$ corresponds to the distribution $\abs{\Psi}^2$, while for subsystems it takes the form $\varsigma[a_i^{A_i}]$, and corresponds to $\abs{\psi_i}^2$. Accordingly, dynamical equilibrium is governed by the progression induced through the gain function, Eq.~\eqref{eqn:gain}—specifically, the gain obtained from deviation which corresponds to feedback terms.
	
	During the interaction process, dynamical equilibrium is initiated at a stable point—assuming that subsystems continuously seek better responses as defined by the gain function. Physically, this means that when subsystems overlap or interact, their pilot waves transition from EWF to CWF, signalling a \emph{shift} in the equilibrium point. Since subsystems are inherently inclined to follow a stable strategy, as their behaviour is governed by the gain function. In this evolution, subsystems continuously refine their strategies through interactions (by the employment of feedbacks, Eq.~\eqref{eqn:time-evo_CWF_one_withaltentfields-explicit} which abide Eq.~\eqref{eqn:gain}) under the guidance of their CWFs, while maintaining a mixed strategy. Over time, they progressively converge toward the best possible responses, thereby naturally preserving the true equilibrium existing at initial and final points while demonstrating a dynamical equilibrium.
	
	While the reign of dynamical equilibrium persists, the continuity equation of the subsystems takes its form, Eq.~\eqref{eqn:con-eq_par-1_fb} (and similarly for Hamilton-Jacobi equation, Eq.~\eqref{eqn:pha-evo_real-par1}), throughout the interval of interactions where feedback terms arise as the difference. These terms, again by the consideration of stable strategy, approximately maintain conservation laws suggested by quantum equilibrium hypothesis—or its conditioned version—throughout the interval while introducing small deviations, leading to oscillations around true equilibrium. As will be shown shortly in a numerical simulation, the combined feedback from all subsystems produces persistent oscillatory behaviour, referencing the true equilibrium form; Eqs.~\eqref{eqn:fb-couple-1st} and~\eqref{eqn:fb-couple-2nd}, ensuring no net gain in the system. As each subsystem strives to approach its conditional quantum equilibrium state, the entire system correspondingly moves toward the quantum equilibrium state.
    
    As a final remark, the concept of dynamical equilibrium presented here prioritizes subsystems and emphasizes the emergent interactions that occur between them. Through the feedback terms, a state of dynamical equilibrium is gradually achieved. In contrast, the causal interpretation~\cite{bohm1954model} prioritizes the entire system, treating it as a conserved fluid that undergoes random fluctuations in both velocity and probability density. These fluctuations tend toward their respective mean values, thereby leading the system toward an equilibrium state—or set of equilibria—defined by the quantum equilibrium hypothesis.
        
    \subsection{Numerical Simulation of a Stable Strategy}
    Consider a two-particle system described by the wave function $\Psi(x_1, x_2, t)$, where $x_1$ and $x_2$ denote the positions of particles 1 and 2 in configuration space, respectively. The CWFs and guiding equations for the subsystems follow from~\cite{doner2022gravitational}. Accordingly, each particle's trajectory is defined by their guidance equation:
    \begin{align}
    	\label{eqn:tra_par-1}X_1(t) &= X_1(0) + \int_0^t \rmd t \frac{\hbar}{m_1} \Im \left(\frac{\nabla \psi_1}{\psi_1}\right)\bigg|_{x = X_1(t)} \\
    	\label{eqn:tra_par-2}X_2(t) &= X_2(0) + \int_0^t \rmd t \frac{\hbar}{m_2} \Im \left(\frac{\nabla \psi_2}{\psi_2}\right)\bigg|_{x = X_2(t)}
    \end{align}
    where $X_1(0)$ and $X_2(0)$ are initially separated, with their positions specified as $X_{1,0} = -a/2$ and $X_{2,0} = a/2$, where $a$ is a constant representing the initial distance between the particles. The potentials for the subsystems are, conditional potential of particle 1 and 2:
    \begin{align*}
    	V(x_1, x_2) \big|_{x_2 = X_2(t)} = V[x, X_2(t)] &= -\frac{G m_1 m_2}{|x - X_2(t)|} \nnl
    	V(x_1, x_2) \big|_{x_1 = X_1(t)} = V[X_1(t), x] &= -\frac{G m_1 m_2}{|X_1(t) - x|}
    \end{align*}	
    Relative conditional potential:
    \begin{align*}
    	V(x_1, x_2) \big|_{x_1 = X_1(t),x_2=X_2(t)} &= V[X_1(t), X_2(t)] \nnl
    	&= -\frac{G m_1 m_2}{|X_1(t) - X_2(t)|}
    \end{align*}
   	It is assumed that the potential for whole system is defined as $V(x_1,x_2,t) = V[x,X_2(t),t] + V[X_1(t),x,t]$ when there is no interaction present (considered as the first scenario). As soon as $V(x_1,x_2,t) \sim \delta(x_1-x_2)$, an interaction emerges between the particles. Consequently, feedback terms start to appear in order of their significance, beginning with the first-order feedback term for particle 1, which is given by:
    \begin{equation}\label{eqn:fb_1st-ord}
    	\frac{\rmd X_2(t)}{\rmd t} \int_0^\tau \rmd t \bigg( \frac{\partial V[x, X_2(t)]}{\partial x_2} - \frac{\partial V[X_1(t), X_2(t)]}{\partial x_2} \bigg) \psi_1(x, t)
    \end{equation}
    where,
    \begin{align*}
    	F_1^{(1)} &= \frac{\partial V[x, X_2(t)]}{\partial x_2} - \frac{\partial V[X_1(t), X_2(t)]}{\partial x_2} \nnl
    	&= G m_1 m_2 \bigg( \frac{x - X_2(t)}{|x - X_2(t)|^3} - \frac{X_1(t) - X_2(t)}{|X_1(t) - X_2(t)|^3} \bigg)
    \end{align*}
    The second-order feedback term for particle 1 is:
    \begin{equation}\label{eqn:fb_2nd-ord}
    	\frac{i\hbar}{2m_2} \int_0^\tau \rmd t \bigg( \frac{\partial^2 V[x, X_2(t)]}{\partial x_2^2} - \frac{\partial^2 V[X_1(t), X_2(t)]}{\partial x_2^2} \bigg) \psi_1(x, t)
    \end{equation}
    where,
    \begin{widetext}
    	\begin{align*}
    		F_1^{(2)} &= \, \frac{\partial^2 V[x, X_2(t)]}{\partial x_2^2} - \frac{\partial^2 V[X_1(t), X_2(t)]}{\partial x_2^2} \nnl
    		&= \, G m_1 m_2 \bigg(\frac{\left(3(x - X_2(t))^3 - |x - X_2(t)|\right)}{|x - X_2(t)|^6} - \frac{\left(3(X_1(t) - X_2(t))^3 - |X_1(t) - X_2(t|)\right)}{|X_1(t) - X_2(t)|^6} \bigg)
    	\end{align*}
    \end{widetext}
    (Accordingly, first-order and second-order feedback terms for particle 2 can be defined in a similar manner.)  Recalling that any non-zero value of $F_1^{(1)}$ or $F_1^{(2)}$ introduces an interaction between the particles, shown by the dynamical evolution of $\psi_1(x,t)$, Eq.~\eqref{eqn:time-evo_CWF_one_withaltentfields-explicit} (with a similar expression holding for particle 2). By utilizing the polar form of the wave functions, as presented in Appendix \ref{app:Class_Feed}, Eqs.~\eqref{eqn:amp-evo_ima-par1} and~\eqref{eqn:pha-evo_real-par1}, where the separate evolutions of the imaginary and real parts were derived (yielding analogous results for particle 2), the result of these interactions become more distinct by first-order feedback term playing its role in the amplitude of the wave function while second-order in the phase of it.
    
    As mentioned, these feedback terms ensure that there is no net gain. Meanwhile, the distinction of these terms by their order, as just discussed, introduces an additional condition. Specifically, for true equilibrium cases, the sum of the first-order feedback terms—representing their coupling—must be equal to zero. The same condition applies to the second-order terms. (While noting that, this coupling creates a feedback loop, a feature unique to two-particle systems: $X_2(t) \rightarrow \psi_1(t,x) \rightarrow X_1(t) \rightarrow \psi_2(t,x) \rightarrow X_2(t)$) Thus, the conditions of the stable strategy correspond to the best-response conditions (as described earlier), expressed as:
    \begin{align}
    	\label{eqn:fb-couple-1st}&\frac{r_1}{\hbar}\frac{dX_2(t)}{\rmd t} \int_0^\tau \rmd t F_1^{(1)} + \frac{r_2}{\hbar}\frac{dX_1(t)}{\rmd t} \int_0^\tau \rmd t F_2^{(1)} = 0\\
    	\label{eqn:fb-couple-2nd}&\frac{\hbar}{2m_2} \int_0^\tau \rmd t F_1^{(2)} + \frac{\hbar}{2m_1} \int_0^\tau \rmd t F_2^{(2)} = 0
    \end{align}
	Assuming that the subsystems are in dynamical equilibrium—functioning as a better-response condition centered on the best-response condition—the conditional and relative conditional potentials are tracked to enable numerical calculation, with gravitational interaction modeled through feedback terms. The separation $|X_2 - X_1|$ is kept non-zero to avoid singularities. The CWFs are assumed to have Gaussian profiles oscillating at given frequencies. Interactions arise from the overlapping of pilot waves, which then influence the particles trajectories as defined by the feedback terms (for particle 1): the first-order term affects the wave function's amplitude, as described in Eq.~\eqref{eqn:amp-evo_ima-par1}, while the second-order term impacts its phase, as shown in Eq.~\eqref{eqn:pha-evo_real-par1}. Now, considering the initial wave function of particle 1 as:
    \begin{equation}
    	\label{eqn:ini_wf-1}\psi_1(t,x) = \big(\frac{1}{2\pi \sigma^2}\big)^{1/4} e^{-\frac{(x-X_{1,0})^2}{2\sigma^2}} \cdot e^{\rmi \phi_1}
    \end{equation}
    and similarly initial wave function of particle 2 as:
    \begin{equation}
    	\label{eqn:ini_wf-2}\psi_2(t,x) = \big(\frac{1}{2\pi \sigma^2}\big)^{1/4} e^{-\frac{(x-X_{2,0})^2}{2\sigma^2}} \cdot e^{\rmi \phi_2}
    \end{equation}
    where $\sigma$ is width of Gaussian wave function, $X_{1,0} = -a/2$ and $X_{2,0} = a/2$ are the initial positions of particle 1 and particle 2, respectively, finally $\phi_1$ and $\phi_2$ are initial phases (due to chosen paths). With these values—where the initial positions of the particles are specified by Eqs.~\eqref{eqn:tra_par-1} and~\eqref{eqn:tra_par-2}, and CWFs are initially random due to their autonomy after subsystems being measured by the magnets, as previously discussed—the particles begin to exhibit behaviour characteristic of dynamical equilibrium, adopting a stable strategy.
    
\begin{figure}
   		\includegraphics[scale=0.49, center]{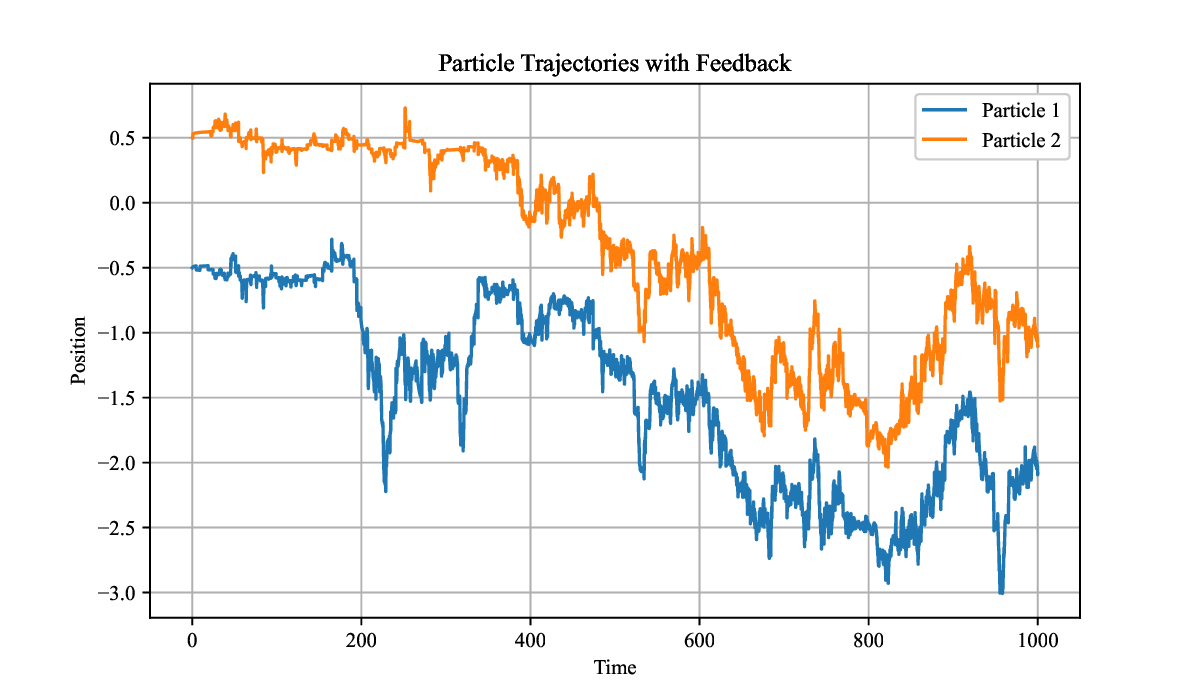}
   		\caption[Numerical calculation of particle interactions]{The plot of numerical calculation based on repeated simulations of the particle interactions through the feedback terms given in Eq.~\eqref{eqn:time-evo_CWF_one_withaltentfields-explicit} where all constants are set to $1$, total simulation time is $1000 \sec$, time steps $\rmd t = 0.01 \sec$, Number of spatial points $N = 1000$ and length of the spatial domain $L = 100$.\label{fig:part-tra_sim}}
   	\end{figure}

   	Numerical calculations based on this initial setup, along with repeated simulations, are presented in Fig~\ref{fig:part-tra_sim}. The feedback terms—Eqs.~\eqref{eqn:fb_1st-ord} and~\eqref{eqn:fb_2nd-ord} (and their counterparts for particle 2)—are the primary contributors to the behaviour depicted, acting through their couplings.
    To observe these couplings in action—as part of the feedback loop mentioned previously—the continuous particle trajectories are discretized into time steps. For this purpose, the approach of Vink~\cite{vink1993quantum} is adopted, wherein he introduced discrete beables and interpreted Bohmian (causal) trajectories as the continuum limit. This discretization naturally introduces transition probabilities, as derived by Vink in order to present the connection of discrete causal trajectories to Nelson's stochastic mechanics~\cite{nelson1966derivation}. In the continuum limit, these probabilities are defined through the relationship between the quantum probability current and the probability density.
    
    In the simulation, while the feedback terms perform their intended function; \emph{measuring} or \emph{monitoring} the other particle, we consider these transition probabilities between time steps as governed by the coupling of the first-order feedback terms, neglecting the remainder of the continuity equation, Eq.~\eqref{eqn:con-eq_par-1_fb}, as it corresponds to true equilibrium cases. Accordingly, we focus directly on Eq.~\eqref{eqn:fb-couple-1st}, with values $=0$, $\geq 0$, and $\leq 0$, and implement these into the particle trajectories. Specifically, when the value is $=0$, particles continue to follow their paths with no change in their positional interval. For $\geq 0$, particles experience an impulse that increases their positional interval, whereas for $\leq 0$, the impulse decreases it. In this context, it is more appropriate to interpret these impulses as changes in the positional interval between particles (increase and decrease), rather than physical direction. Meanwhile, the coupling of the second-order feedback terms, Eq.~\eqref{eqn:fb-couple-2nd}, determines the magnitude of these changes. (To clarify, refer to Eq.~\eqref{eqn:pha-evo_real-par1} while considering $\frac{\partial p_1}{\partial t} = \frac{\partial}{\partial x} (\frac{\partial s_1}{\partial t})$ relation, which shows the dependence of each particle's momentum on the other.) 
    
    The analysis of the simulation, in Fig~\ref{fig:part-tra_sim}, shows that, starting from the time intervals $300$–$350$, the feedback terms visibly begin to effect the particle trajectories. Through the mechanism described above, the particles exhibit a syncing behaviour, where each deviation is compensated by the action of dynamical equilibrium, thereby maintaining a stable average distance between them throughout their evolution. As the number of time steps increases, similar occurrences continue to arise, with the compensatory mechanism of dynamical equilibrium remaining consistently active.
    
    An important point should be addressed here: these numerical simulations, which consider a two-particle system—where each particle's environment is limited to the other particle—represent a Non-Markovian process (again discussed as a loop mechanism) and, as a result, differ from Bohm and Vigier’s causal interpretation scheme~\cite{bohm1954model}. Consequently, the results obtained here should be interpreted with caution, considering them only as a framework useful for illustrating a simplified form of interactions between subsystems under specific conditions. Therefore, aside from the initial randomness in particle velocities (again comes from CWF's brief autonomy)-deliberately restricted to their directions and used solely to determine the order in which one particle influences the other-this scheme lacks any form of stochastic behaviour.(This deliberate choice ensures that the purest form of dynamical equilibrium governs the subsystem dynamics, resembling the idea of superdeterminism~\cite{hossenfelder2020rethinking})
    
    However, the Non-Markovian process changes as soon as $N$-particle systems are considered, which is the case for both the quantum equilibrium hypothesis and its conditional version. Naturally, this consideration paves the way for future studies as the next logical step. Final remark, this simulation is intended solely to provide a clearer understanding of dynamical equilibrium by visualizing it through particle trajectories and should not be interpreted as a definitive result.
    
    \section{Conclusion}
    In this work, two essential topics were addressed. The first involved an assessment of the model presented in \cite{doner2022gravitational}, demonstrating that the generation (or increase) of entanglement from weak entanglement—in the absence of entanglement potential fields—is possible. To achieve this, feedback terms derived from gravitational potential interactions are employed, as shown in Eq.~\eqref{eqn:time-evo_CWF_one_withaltentfields-explicit}. These feedback terms correspond to the overlapping of pilot waves (CWFs) in physical space. 
    
    The next step followed the standard procedures of the model, leading to the witness function, $W$. This function now incorporates the phases of weak entanglement and appropriately uses chosen flight times—shortened from the initial time—to prevent the emergence of entanglement potential fields. In the end, the assessment yielded two key results: first, that entanglement can indeed emerge from weak entanglement; and second, that weak entanglement is the underlying cause of entanglement potential fields, as expressed in Eq.~\eqref{eqn:time-evo_CWF_one_withaltentfields-explicit}.
    
    The second topic builds on the first by exploring these feedback terms as a potential communication scheme. It is important to clarify that, although these terms arise from classical potentials, the communication scheme in which they are employed is not classical and therefore cannot be classified as a local operations and classical communication (LOCC) channel. In this context, the classical potentials do not function as direct mediators of communication; rather, they serve as sources along Bohmian trajectories, consistent with their role in the original model~\cite{doner2022gravitational}.
    
    From this communication framework, a dynamical version of the conditional quantum equilibrium hypothesis—referred to as dynamical equilibrium—was developed. This equilibrium accounts for the continuous interactions between the subsystems and differs from the quantum equilibrium hypothesis and its conditioned version, in which the typical distribution condition must hold at all times. The key distinction is that, rather than assuming equilibrium is always maintained, the system is dynamically driven toward it. To formalize this behaviour, the feedback terms in Eq.~\eqref{eqn:time-evo_CWF_one_withaltentfields-explicit} were applied within a stable strategy framework, as defined by Nash's Theorem.
    
    As a final remark—and for the sake of future studies on dynamical equilibrium—the classification of the continuous interactions it comprises is essential. Following the analysis in~\cite{jacobs2006straightforward}, it becomes clear that these interactions fall under the category of continuous weak measurements, as they occur over infinitesimally small time steps, $\Delta t \rightarrow 0$ or $\Delta t \rightarrow \rmd t$. (By contrast, the quantum equilibrium hypothesis and its conditional version correspond to strong measurements.) It is important to note that the numerical simulations demonstrated here represent a discrete approximation of this process. This distinction establishes a link between dynamical equilibrium and the framework of continuous weak measurement in quantum measurement theory~\cite{patel2017weak}. 
    
    A natural extension of this idea—where a single particle interacts with an $N$-particle environment—defines a promising direction for future research. This involves applying the feedback scheme through particle trajectories, as in dynamical equilibrium, but generalized to systems with many particles. However, this generalization inherently disrupts the closed feedback loop structure that is characteristic of two-particle systems. Therefore, a more refined approach is required—one that simultaneously accounts for both the particle's interactions with the $N$-particle environment and the internal interactions within the environment itself at each infinitesimal time step—in order to effectively extend the framework.
    
    Another promising direction concerns the possibility of non-equilibrium in the quantum equilibrium hypothesis and its conditional version, which may be even more relevant in the setting of dynamical equilibrium~\cite{valentini1991signal1}. In particular, the relaxation properties of such non-equilibrium states—already proposed as potentially experimentally distinguishable~\cite{colin2010quantum}—warrant further investigation.          
    \appendix
    
    \section{Nash's Theorem - A Stable Strategy}\label{app:Nash-Th}
    Dynamical equilibrium is defined as a stable strategy for two agents (particles) based on Nash's Theorem~\cite{nash1950equilibrium}, which states that every game with a finite number of agents and action profiles has at least one Nash equilibrium~\cite{jiang2009tutorial}. A Nash equilibrium is defined a stable strategy profile where no agent benefits from unilateral deviation. The existence of such an equilibrium is guaranteed by a function satisfying Brouwer’s fixed-point theorem (a generalization of Sperner's Lemma, which states that any continuous function mapping a compact, convex subset of Euclidean space to itself must have a fixed point~\cite{jiang2009tutorial}), which ensures at least one fixed point. This function operates within the space of mixed strategies—probability distributions over possible actions. The collection of all such distributions forms a simplex:
    \begin{equation*}
    	\Delta^n = \{\vec{x} \in \mathbb{R}^{n+1} \mid \sum_{i=1}^{n+1} \vec{x}_i = 1, \vec{x}_i \geq 0\}
    \end{equation*}
    For multiple players, mixed strategy profiles span the Cartesian product of simplices, forming a simplotope (for simplicity the Cartesian product of two simplices given as: $\Delta^1 \times \Delta^1 = \{ (\vec x, \vec y) \in \mathbb{R}^2 | 0 \leq \vec x \leq 1 , 0 \leq \vec y \leq 1 \}$ where it describes a square in $\mathbb{R}^2$). A function mapping strategy profiles to themselves identifies Nash equilibria as fixed points. A natural approach is the best-response function, but discontinuities arise due to multiple possible best responses. Instead, a better-response condition is introduced, ensuring a unique, continuous mapping. This function updates probabilities incrementally, and evidently avoid abrupt shifts. For that the advantage function is defined:
    \begin{equation}
    	\text{Adv}(\varsigma) = \varsigma'
    \end{equation}
    each component of the output vector $\varsigma'$ is given by:
    \begin{equation}\label{eqn:each-com}
    	\varsigma'[a_i^j] = \frac{\varsigma[a_i^j] + \text{Gain}(\varsigma, a_i^j)}{\sum_{k=1}^{A_i} \varsigma[a_i^k] + \text{Gain}(\varsigma, a_i^k)}
    \end{equation}
    which adjusts probabilities based on the expected gain of deviating from a given strategy: 
    \begin{equation}\label{eqn:gain}
    	\text{Gain} (\varsigma, a_i^j) = \text{max} \{0, \nu_i(a_i^j, \varsigma) - \nu_i(\varsigma) \}
    \end{equation}
    where $\nu_i(a_i^j, \varsigma)$ represents the expected utility of action $a_i^j$. The updated probability distribution is normalized to maintain a valid mixed strategy. Since the advantage function is continuous and maps a convex, compact space to itself, Brouwer’s fixed-point theorem guarantees at least one fixed point. At equilibrium, all gains are zero, ensuring no player has an incentive to deviate. Noting that, in the context of causal interpretation this result addressed by the equality of maximum and minimum limits where they approach to a constant limit and evidently proves that $\mathbb{P}(t,x) \rightarrow \rho_0(t,x)$. Thus, every finite game has at least one Nash equilibrium, serving as a reference point for dynamical evolution.
    %The final task is to identify these Nash equilibria, which will serve as reference points for the dynamic evolution of better-response functions. To proceed, we now define the best-response condition.    
    \subsection{Example: Nash equilibria for two agents, defining-best response condition as the reference:}\label{app:best-response}
    Consider an $m \times n$ bipartite game with two players, where $A$ and $B$ are $m \times n$ payoff matrices. The entry $a_{ij}$ in $A$ represents player 1's payoff when choosing row $i$ while player 2 chooses column $j$, and $b_{ij}$ in $B$ represents player 2's payoff under the same conditions. These matrices capture the expected utilities ($\nu_i$), assumed to be nonnegative. A mixed strategy for player 1 is a probability distribution over $m$ rows, represented by an $m$-vector $x \in \mathbb{R}^m$, satisfying $x \geq \vec{0}$ and $\vec{1}^\top x = 1$. Similarly, a mixed strategy for player 2 is an $n$-vector $y \in \mathbb{R}^n$, satisfying $y \geq \vec{0}$ and $\vec{1}^\top y = 1$. This defines the mixed strategy sets:
    \begin{align*}
    	X &= \{ x \in \mathbb{R}^m \mid x \geq \vec{0}, \vec{1}^\top x = 1 \}, \nnl
    	Y &= \{ y \in \mathbb{R}^n \mid y \geq \vec{0}, \vec{1}^\top y = 1 \}
    \end{align*}
    A best response for player 1 to player 2's strategy $y$ is an $x \in X$ that maximizes the expected payoff $x^\top A y$. Similarly, a best response for player 2 to $x$ is a $y \in Y$ that maximizes $x^\top B y$. A Nash equilibrium is a pair $(x, y) \in X \times Y$ where $x$ and $y$ are best responses to each other.
    
    A mixed strategy $x$ is a best response to $y$ if and only if all pure strategies in its support are pure best responses to $y$. That is, for all $i = 1, \dots, m$:
    \begin{equation}
    	x_i > 0 \Rightarrow (Ay)_i = \nu = \text{max} \{ (Ay)_k | k = 1,...,m \}
    \end{equation}
    where $(Ay)_i$ represents the expected payoff to player 1 for row $i$. This implies:
    \begin{align}
    	x^\top Ay = \sum_{i=1}^m x_i(Ay)_i &= \sum_{i=1}^m x_i(\nu - (\nu - (Ay)_i) \nnl
    	&= u - \sum_{i=1}^m x_i (\nu - (Ay)_i
    \end{align}
    Since $x_i \geq 0$ and $\nu - (Ay)_i \geq 0$ for all $i$, it follows that $x^\top Ay \leq \nu$, with equality ($x^\top Ay = \nu$) holding if and only if $x_i > 0 \Rightarrow (Ay)_i = \nu$. By applying the best-response condition, the problem reduces from an infinite set of mixed strategies to a finite condition involving only pure strategies, simplifying the calculations.
        
    \section{Classification of Feedback Terms by Their Order}\label{app:Class_Feed}
    
    Before presenting numerical simulations, it is useful to classify the feedback terms by their order. If such a classification is possible, it can lead to a neat and convenient approach: a coupling where the first-order feedback term affecting particle 1 is paired with its counterpart affecting particle 2, Eq.~\eqref{eqn:fb-couple-1st}. A similar result can be established for second-order feedback terms, Eq.~\eqref{eqn:fb-couple-2nd}. Interestingly, this result can be obtained quite naturally by considering the two-particle system’s wave function in polar form, $\Psi(t,x_1,x_2)) = R(t,x_1,x_2) \exp{\rmi S(t,x_1,x_2)/\hbar}$, along with the CWF for each particle follow as:
    \begin{align*}	
		\psi_1(t,x)
    	% &= R(t,x,X_2(t))\exp{S(t,x,X_2(t))/\hbar} \nnl
    	&= r_1(t,x,X_2(t)) e^{\rmi s_1(t,x,X_2(t))/\hbar} \nnl
    	\psi_2(t,x) 
    	% &= R(t,X_1(t),x)\exp{S(t,X_1(t),x)/\hbar} \nnl
    	&= r_2(t,X_1(t),x) e^{\rmi s_2(t,X_1(t),x)/\hbar}
    \end{align*}
    %(where the subscripts of $r$ and $s$ to $x$, similarly for particle 2);
    %\begin{equation*}
    	%\begin{split}
    		%r_1(t,x,X_2(t)) &= R(t,x,x_2)|_{x_2=X_2(t)}\\
    		%r_2(t,X_1(t),x) &= R(t,x_1,x)|_{x_1=X_1(t)}\\
    		%s_1(t,x,X_2(t)) &= S(t,x,x_2)|_{x_2=X_2(t)}\\
    		%s_2(t,X_1(t),x) &= S(t,x_1,x)|_{x_1=X_1(t)}
    	%\end{split}
    %\end{equation*}
    L.h.s. of \schr~equation, Eq.~\eqref{eqn:time-evo_CWF_one_withaltentfields-explicit}:
    \begin{equation*}
    	\rmi\hbar \frac{\partial \psi_1(t,x)}{\partial t}
    	%\rmi\hbar \frac{\partial}{\partial t} \Big( r_1(t,x,X_2(t)) e^{\rmi s_1/\hbar} \Big)
    	= \bigg( \rmi \hbar \frac{\partial r_1}{\partial t} - r_1 \frac{\partial s_1}{\partial t} \bigg) e^{\rmi s_1/\hbar}
    \end{equation*}
\iffalse
    Following this, with the first term on the r.h.s. of Eq.~\eqref{eqn:time-evo_CWF_one_withaltentfields-explicit}:
    \begin{widetext}
    	\begin{equation*}
    		\begin{split}
    			-\frac{\hbar^2}{2m} \frac{\partial^2 \psi_1(t,x)}{\partial x^2}
    			% &-\frac{\hbar^2}{2m_1} \frac{\partial}{\partial x}\bigg( \frac{\partial r_1}{\partial x} e^{\rmi s_1/\hbar} + \frac{\rmi r_1}{\hbar} \frac{\partial s_1}{\partial x} e^{\rmi s_1/\hbar} \bigg)\\
    			% &=- \frac{\hbar^2}{2m_1} \bigg( \frac{\partial^2 r_1}{\partial x^2} e^{\rmi s_1/\hbar} + \frac{2\rmi r_1}{\hbar} \frac{\partial s_1}{\partial x} e^{\rmi s_1/\hbar} + \frac{\rmi r_1}{\hbar} \frac{\partial^2 s_1}{\partial x^2} e^{\rmi s_1/\hbar} - \frac{r_1}{\hbar^2} \Big( \frac{\partial s_1}{\partial x} \Big)^2 e^{\rmi s_1/\hbar} \bigg)\\
    			&=- \frac{\hbar^2}{2m_1} \bigg( \frac{\partial^2 r_1}{\partial x^2} + \frac{2\rmi}{\hbar} \frac{\partial r_1}{\partial x} \frac{\partial s_1}{\partial x} + \frac{\rmi r_1}{\hbar} \frac{\partial^2 s_1}{\partial x^2} + \frac{r_1}{\hbar^2} \Big( \frac{\partial s_1}{\partial x} \Big)^2 \bigg) e^{\rmi s_1/\hbar}
    		\end{split}
    	\end{equation*}
    \end{widetext}
\fi
    The \schr~equation of by Eq.\eqref{eqn:time-evo_CWF_one_withaltentfields-explicit} particle 1 follows, with neglecting $f(t)\psi_1$, while $e^{\rmi s_1/\hbar}$ terms cancel each other (where $\tau$ is the time interval that entails period of CWF, it is in the $\Gamma$ in our calculations before):
    \begin{widetext}
    	\begin{equation}
    		\begin{split}
    			\left( \rmi \hbar \frac{\partial r_1}{\partial t} - r_1 \frac{\partial s_1}{\partial t} \right) = 		&-\frac{\hbar^2}{2m_1} \left( \frac{\partial^2 r_1}{\partial x^2} + 2 \frac{\rmi}{\hbar} \frac{\partial r_1}{\partial x} \frac{\partial s_1}{\partial x} - \frac{r_1}{\hbar^2} \left( \frac{\partial s_1}{\partial x} \right)^2 + \frac{\rmi r_1}{\hbar} \frac{\partial^2 s_1}{\partial x^2} \right) + V[t,x,X_2(t)] r_1 \\
    			&+ \frac{\rmd X_2(t)}{\rmd t} \int_0^\tau \rmd t \left( \frac{\partial V}{\partial x_2}[t,x,X_2(t)] - \frac{\partial V}{\partial x_2}[t,X_1(t),X_2(t)] \right) r_1 \\
    			& - \frac{\hbar}{2m_2} \int_0^\tau \rmd t \left( \frac{\partial^2 V}{\partial x_2^2}[t,x,X_2(t)] - \frac{\partial^2 V}{\partial x_2^2}[t,X_1(t),X_2(t)] \right) r_1
    		\end{split}        
    	\end{equation}
    \end{widetext}
	Accordingly, imaginary part follows:
    \begin{widetext}
    	\begin{equation}\label{eqn:amp-evo_ima-par1}
    		\frac{\partial r_1}{\partial t} = \frac{1}{m_1} \frac{\partial r_1}{\partial x} \frac{\partial s_1}{\partial x} + \frac{r_1}{2m_1} \frac{\partial^2 s_1}{\partial x^2}
    		+ \frac{r_1}{\hbar}\frac{dX_2(t)}{\rmd t} \int_0^\tau \rmd t \left( \frac{\partial V}{\partial x_2}[t,x,X_2(t)] - \frac{\partial V}{\partial x_2}[t,X_1(t),X_2(t)] \right)
    	\end{equation}
    \end{widetext}
    To derive the continuity equation for a single particle, we use the relation $\partial r^2_1/\partial t = 2r_1 \partial r_1/\partial t$, which results:
    \begin{widetext}
    	\begin{equation}\label{eqn:con-eq_par-1_fb}
    		\frac{\partial r^2_1}{\partial t} =  \frac{1}{m_1}\frac{\partial}{\partial x} \bigg( r^2_1 \frac{\partial s_1}{\partial x} \bigg) 
    		+ \frac{2r_1^2}{\hbar}\frac{dX_2(t)}{\rmd t} \int_0^\tau \rmd t \left( \frac{\partial V}{\partial x_2}[t,x,X_2(t)] - \frac{\partial V}{\partial x_2}[t,X_1(t),X_2(t)] \right)
    	\end{equation}
    \end{widetext}
    The first-order feedback term involves the gradient of the conditional potential minus the gradient of the relative conditional potential. This term represents the feedback \emph{impulse} on the single-particle's pilot wave function, guiding it back to its equilibrium state. The real part:
    \begin{widetext}
    	\begin{equation}\label{eqn:pha-evo_real-par1}
    		\frac{\partial s_1}{\partial t} = - \frac{1}{2m_1} \left( \frac{\partial s_1}{\partial x} \right)^2 + V[t,x,X_2(t)] - V_\text{qu}^{\psi_1}(t,x)\\ 
    		- \frac{\hbar}{2m_2} \int_0^\tau \rmd t \left( \frac{\partial^2 V}{\partial x_2^2}[t,x,X_2(t)] - \frac{\partial^2 V}{\partial x_2^2}[t,X_1(t),X_2(t)] \right)
    	\end{equation}
    \end{widetext}
    where $V_\text{qu}^{\psi_1} = - \frac{\hbar^2}{2m_1r_1} \frac{\partial^2 r_1}{\partial x^2}$ is polar form quantum potential of particle 1. By moving all the terms on the r.h.s. of the equation to the l.h.s., the Hamilton-Jacobi equation for particle 1 can be readily derived. Thus, it can be observed that the second-order feedback term—defined as the second-order gradient of the conditional potential minus the second-order gradient of the relative conditional potential—represents the feedback of \emph{pressure} on the system.
    
    In the absence of interaction, these equations lose the feedback terms, meaning that the pilot waves of the particles do not overlap, and consequently, no entanglement is created. As a result, the last term of the Hamilton-Jacobi equation does not emerge. Additionally, if there is no initial entanglement leaving only the conditional potential term. This term then becomes equal to $V_1(t,x_1)$ (as discussed in the reference situation).
    
    The derivations so far can be summarized by the evolution of the imaginary part of the wave function for particle 1, given by Eq.~\eqref{eqn:amp-evo_ima-par1}, and the real part, given by Eq.~\eqref{eqn:pha-evo_real-par1}. Similarly, these derivations apply to particle 2.
    
    Accordingly, the feedback terms are active only during the interaction period, denoted by the time interval $\tau$, as shown in Eqs.~\eqref{eqn:amp-evo_ima-par1} and~\eqref{eqn:pha-evo_real-par1}, where the pilot waves are described as CWFs. Conversely, when the pilot waves are described as EWFs—indicating the absence of interactions (in the fapp sense)—these terms vanish.
    
    Since each particle follows a stable strategy—meaning that each responds with awareness of the other’s behaviour and consistently aims to align with a best-response condition—zero gain must be achieved. This requirement implies that the feedback terms, as given in Eqs.~\eqref{eqn:amp-evo_ima-par1} and~\eqref{eqn:pha-evo_real-par1}, must continuously sum to zero over the time interval $\tau$, which defines the duration of the dynamical equilibrium. Accordingly, the coupling of feedback terms according to their order reflects the stable strategies adopted by the subsystems.
    
    \newpage
  
    \bibliographystyle{unsrt}
    \bibliography{references}
\end{document}